\documentclass[journal]{IEEEtran}
\usepackage{amsmath,amsfonts,amssymb}
\usepackage{algorithmic}
\usepackage{algorithm}
\usepackage{array}
\usepackage[caption=false,font=footnotesize]{subfig}
\usepackage{textcomp}
\usepackage{stfloats}
\usepackage[hyphens]{url}
\usepackage{graphicx}
\usepackage{booktabs}
\usepackage{cite}
\usepackage{color}
\usepackage{multirow}

\graphicspath{{./figures/}}
\usepackage[hidelinks]{hyperref}

\begin{document}

\title{Interpretable Reinforcement Learning for Industrial Park Energy Management with Heterogeneous Batteries under Demand Response}

\author{Meng Yuan,~\IEEEmembership{Senior Member,~IEEE,} Tinghui Yan, and Zhezhuang Xu,~\IEEEmembership{Member,~IEEE}
\thanks{Meng Yuan is with the School of Engineering and Computer Science, Victoria University of Wellington, Kelburn 6012, New Zealand (e-mail: meng.yuan@vuw.ac.nz).}\thanks{Tinghui Yan and Zhezhuang Xu are with the College of Electrical Engineering and Automation, Fuzhou University, Fuzhou 350108, China (e-mail: zzxu@fzu.edu.cn).}}

\markboth{Preprint}{Yuan \MakeLowercase{\textit{et al.}}: Interpretable RL for Industrial Park Energy Management}

\maketitle

\begin{abstract}
Industrial park dispatch must balance demand response (DR) revenue with office comfort, electric vehicle (EV) departure requirements and battery ageing. Neural policies complicate constraint enforcement and operator inspection. In this work, we propose an interpretable constrained deep reinforcement learning (DRL) method for industrial parks with photovoltaic generation and heterogeneous batteries. A deep deterministic policy gradient (DDPG) agent with dual-ascent multipliers, termed DDPG-Lag, optimises net operating profit with chemistry-specific ageing costs. Its policy is distilled into shallow decision trees using physics-informed features. A residual tree controls heating, ventilation and air conditioning through a model-predicted comfort projection. Battery dispatch is determined by DDPG-Lag, and its trees provide readable policy summaries. Case studies with outdoor temperature forecast errors and uncertain EV departures show that the proposed method achieves 0.04~h of average comfort violations and 98\% departure compliance, with a 9.5\% cost reduction relative to a rule-based strategy. Relative to fixed-penalty DRL, comfort and departure violations are reduced by 99.9\% and 72.7\% to 95.4\%, respectively.
\end{abstract}


\section{Introduction}

\IEEEPARstart{I}{ndustrial} parks increasingly combine production lines and offices with rooftop photovoltaic (PV) systems, stationary energy storage systems (ESSs) and electric vehicles (EVs) parked on site. ESSs store surplus PV energy and release it during supply shortfalls, which raises self-consumption and supply resilience~\cite{datta2021review,hannan2021battery}. Through vehicle-to-grid operation, EVs add mobile storage, and the coordinated dispatch of ESSs and EVs is used in building energy management~\cite{yilmaz2012review,zhang2024multi,liu2026deep}. Demand response (DR) programmes reward users who reduce their consumption during grid peak periods~\cite{siano2014demand}, and energy management with DR can lower both the operating cost and the emissions of local energy systems~\cite{dey2023microgrid}. Because production loads can rarely be curtailed, an industrial park can instead discharge its batteries during DR events to lower its metered consumption.

Office zones must maintain thermal comfort, which limits the flexibility of their heating, ventilation and air conditioning (HVAC) loads~\cite{tang2019model}, whereas production zones pursue low operating costs and emissions. In two case studies, aggregating the loads of multiple buildings reduced the peak demand by about 5\% to 6\% relative to the sum of the individual peaks~\cite{2004Benefits}. Coordinated control of multiple buildings during DR events also causes less occupant discomfort than controlling each building individually~\cite{2017Conceptual}. Moreover, each battery cycle causes ageing, and stationary and EV batteries have heterogeneous characteristics that call for separate models~\cite{yuan2026reinforcement}. Building energy management systems (EMSs) support energy monitoring~\cite{selvaraj2023smart} and coordinated resource scheduling~\cite{thirugnanam2022energy}.

Model predictive control (MPC) and meta-heuristic optimisation are common choices for building EMSs~\cite{schmitt2023regression,khan2022modeling}. Dispatch models balance physical detail and computational cost~\cite{kuboth2019economic}. Reinforcement learning (RL) learns a dispatch policy from interaction data. Deep RL (DRL) has been applied to battery arbitrage with ageing models~\cite{cao2020deep} and to buildings with heterogeneous batteries~\cite{liu2026deep}. Multi-agent RL has also been used for DR pricing and incentives~\cite{2024Demand}. Hierarchical multi-agent learning has coordinated HVAC comfort control with distributed energy resources~\cite{pei2025hierarchical}. Manually weighted reward penalties require tuning and can allow constraint violations~\cite{su2025review}. Constrained formulations treat operating requirements as explicit constraints~\cite{altman1999constrained}, and constrained policy optimisation~\cite{achiam2017constrained} and Lagrangian methods~\cite{tessler2019reward} incorporate them into policy optimisation.

Neural policies are black boxes, which makes them difficult for operators to understand and trust~\cite{chen2023interpretable}. For high-stakes decisions, inherently interpretable models are preferable to post-hoc explanations when they perform comparably~\cite{rudin2019stop}. Policy extraction imitates a neural policy with an interpretable model such as a decision tree (DT)~\cite{bastani2018verifiable}, and decision rules have been extracted from DRL energy management policies of building complexes~\cite{chen2025novel}. For industrial parks, readable rules must encode the energy required before EV departure and the HVAC power needed to maintain comfort. Physics-informed features express these requirements directly and support executable residual HVAC control.

To address these issues, this article proposes an interpretable constrained DRL method for the energy management of an industrial park with PV, an ESS, EVs and production and office loads under a tiered DR scheme with uncertain outdoor temperature forecasts and EV departure times. The main contributions of this work are summarised as follows:
\begin{enumerate}
\item A DR-based energy management formulation for an industrial park with heterogeneous loads and batteries, in which a dynamic distribution ratio decides how much battery discharge is credited to the metered production load, with dispatch-level ageing costs and explicit comfort and EV departure constraints.
\item A deep deterministic policy gradient (DDPG) agent with an augmented Lagrangian reward, termed DDPG-Lag, in which the observed comfort and departure violations drive dual-ascent updates of the penalty weights and select the final policy on the training data.
\item A physics-informed distillation of DDPG-Lag into 4 shallow regression trees, in which a residual on the required HVAC power, projected onto the model-predicted comfort band, forms an executable HVAC rule and the battery trees summarise the learned dispatch in readable rules.
\end{enumerate}

The remainder of this article is organised as follows. Section~\ref{sec:Problem_formulation} presents the system models and the problem formulation. Section~\ref{sec:Proposed method} details the proposed method. Section~\ref{sec:Results} presents the case studies, and Section~\ref{sec:Conclusion} concludes the article.

\section{System Models and Problem Formulation}\label{sec:Problem_formulation}

\begin{figure}[tb]
\centering
\includegraphics[width=\columnwidth]{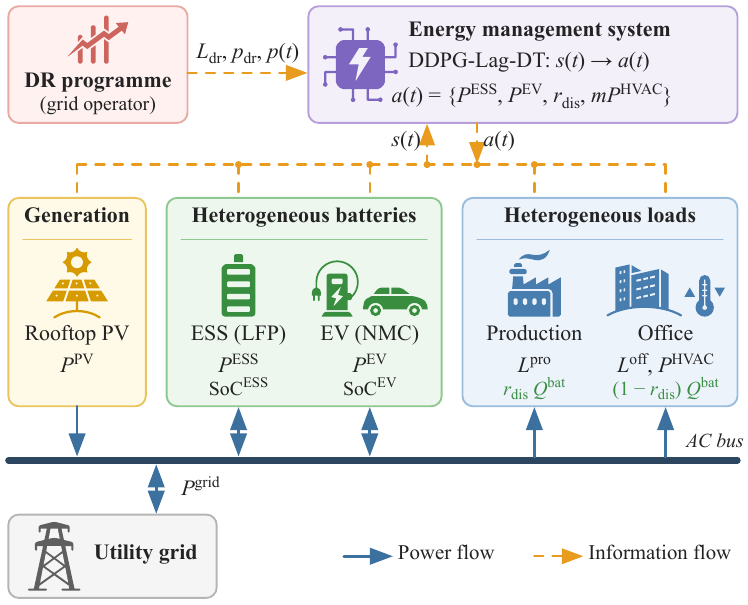}
\caption{Industrial park energy management system.}
\label{fig:schematic}
\end{figure}

The industrial park considered in this study is shown in Fig.~\ref{fig:schematic}. It comprises a production zone, an office zone, rooftop PV, a stationary ESS and a fleet of EVs that is parked on site during working hours. The EMS dispatches the 2 batteries and the HVAC system of the office zone, and it allocates the battery energy between the 2 zones. The battery energy allocated to each zone is indicated by the green labels. The park operates in discrete time slots $t \in \{1, \ldots, T_{\mathrm{total}}\}$ of length $\Delta t$.

\subsection{System Modelling}

\subsubsection{HVAC Model}
The indoor air temperature $T_{\mathrm{in}}(t)$ serves as the comfort proxy of the office zone, and the thermal dynamics of the building follow a first-order model~\cite{lu2020optimal}:
\begin{equation}
\begin{split}
T_{\mathrm{in}}(t+1) &= \varepsilon T_{\mathrm{in}}(t) + (1-\varepsilon)T_{\mathrm{out}}(t)\\
&\quad + (1-\varepsilon)m(t)\frac{\eta^{\mathrm{HVAC}} P^{\mathrm{HVAC}}(t)}{G_{\mathrm{b}}},
\end{split}
\label{eq:temperature}
\end{equation}
where $T_{\mathrm{out}}(t)$ is the outdoor temperature, $\varepsilon$ is the thermal inertia factor, $G_{\mathrm{b}}$ is the thermal conductance of the building, $\eta^{\mathrm{HVAC}}$ is the efficiency ratio, $P^{\mathrm{HVAC}}(t) \ge 0$ is the HVAC power and $m(t) \in \{1,-1,0\}$ selects heating, cooling or the off mode. The HVAC power and the indoor temperature are limited as follows:
\begin{equation}
0 \le P^{\mathrm{HVAC}}(t) \le P^{\mathrm{HVAC}}_{\max}, \quad T_{\min} \le T_{\mathrm{in}}(t) \le T_{\max},
\label{eq:comfort}
\end{equation}
where $P^{\mathrm{HVAC}}_{\max}$ is the rated HVAC power and $T_{\min}$ and $T_{\max}$ are the comfort bounds.

\subsubsection{Battery Model}
Stationary storage and EVs tend to use different lithium-ion chemistries. In this study, the ESS uses lithium iron phosphate (LFP) cells, which offer higher safety and longer cycle life, and the EVs use lithium nickel manganese cobalt oxide (NMC) cells, which provide higher energy density but age faster. The state of charge (SoC) of battery $j \in \{\mathrm{ESS},\mathrm{EV}\}$ is denoted by $\text{SoC}^j(t)$. Both batteries share the same energy model~\cite{cao2020deep}:
\begin{equation}
\begin{aligned}
&\text{SoC}^{j}(t+1) =\\
&\begin{cases}
\text{SoC}^{j}(t) + \dfrac{\eta^j_{\mathrm{c}}P^{j}(t)\Delta t}{E^{j}}, & P^{j}(t) > 0, \\
\text{SoC}^{j}(t) + \dfrac{P^{j}(t)\Delta t}{\eta^j_{\mathrm{d}}E^{j}}, & P^{j}(t) < 0, \\
\max\left\{\text{SoC}^j_{\min},\,\text{SoC}^{j}(t)-\frac{P_{\mathrm{sb}}^{j}\Delta t}{E^{j}}\right\}, & P^{j}(t) = 0,
\end{cases}
\end{aligned}
\label{eq:soc}
\end{equation}
where $P^{j}(t)$ is the battery power, positive when charging, $E^{j}$ is the capacity, $\eta^j_{\mathrm{c}}$ and $\eta^j_{\mathrm{d}}$ are the charging and discharging efficiencies, $P^{j}_{\mathrm{sb}}$ is the standby power loss of an idle battery, saturated at the lower SoC bound, and the binary variables $s^j_{\mathrm{c}}(t)$ and $s^j_{\mathrm{d}}(t)$ indicate charging and discharging according to the sign of $P^j(t)$. The SoC, the power and the operating mode are constrained as follows:
\begin{equation}
\begin{gathered}
\text{SoC}^j_{\min} \le \text{SoC}^{j}(t) \le \text{SoC}^j_{\max},\\
|P^{j}(t)| \le P^{j}_{\mathrm{rated}}, \quad s^j_{\mathrm{d}}(t) s^j_{\mathrm{c}}(t) = 0,
\end{gathered}
\label{eq:battery_limits}
\end{equation}
where $\text{SoC}^j_{\min}$ and $\text{SoC}^j_{\max}$ are the SoC bounds and $P^{j}_{\mathrm{rated}}$ is the rated power. The EVs are modelled as an aggregated battery that is available only while the vehicles are on site. Let $e(t) \in \{0,1\}$ indicate their presence. The EV fleet arrives with $\text{SoC}_{\mathrm{arr}}$ and draws no charging or discharging power while absent. A departure occurs at time slot $t$ when $e(t) = 1$ and $e(t+1) = 0$, and the EVs must then hold enough energy for the next trip:
\begin{equation}
\text{SoC}^{\mathrm{EV}}(t+1) \ge \text{SoC}_{\mathrm{lim}}, \quad t \in \mathcal{T}_{\mathrm{dep}},
\label{eq:departure}
\end{equation}
where $\mathcal{T}_{\mathrm{dep}}$ is the set of departure time slots and $\text{SoC}_{\mathrm{lim}}$ is the minimum departure SoC. The EMS uses the reported departure schedule for planning, while actual departures follow the observed plug status.

\subsubsection{Power Balance}
The grid supplies the remaining demand of the park:
\begin{equation}
\begin{split}
P^{\mathrm{grid}}(t) + P^{\mathrm{PV}}(t) &= L^{\mathrm{pro}}(t) + L^{\mathrm{off}}(t) + P^{\mathrm{HVAC}}(t)\\
&\quad + P^{\mathrm{ESS}}(t) + P^{\mathrm{EV}}(t),
\end{split}
\label{powerblance}
\end{equation}
where $P^{\mathrm{PV}}(t)$ is the PV output, which is not controlled, $L^{\mathrm{pro}}(t)$ and $L^{\mathrm{off}}(t)$ are the production and office loads without the HVAC system, and $P^{\mathrm{grid}}(t) > 0$ denotes energy purchased from the grid.

\subsection{Problem Formulation}

The EMS maximises the expected net operating profit, which is the DR revenue minus the grid, carbon and battery ageing costs, subject to the system model:
\begin{equation}
\begin{split}
\max\ &\mathbb{E}\Bigl[\sum_{t=1}^{T_{\mathrm{total}}} \bigl(R_{\mathrm{dr}}(t) - C_{\mathrm{grid}}(t) - C_{\mathrm{CO_2}}(t) - C_{\mathrm{deg}}(t)\bigr)\Bigr]\\
\text{s.t.}\ &\eqref{eq:temperature}\text{--}\eqref{powerblance},
\end{split}
\label{eq:problem}
\end{equation}
where the expectation is taken over the uncertain PV output, loads, outdoor temperature and EV departure times. The negative of the profit is reported as the net operating cost. Each term is defined in the following.

\subsubsection{DR Revenue}
The DR revenue is evaluated hourly using a tiered compensation scheme with a baseline load~\cite{fujian2024demandresponse}. The grid operator invites participants a day ahead to reduce their consumption during peak hours, and the response is measured against the baseline. Let $H$ denote the number of hours per day. The baseline at intra-day hour $h \in \{1,\ldots,H\}$ is the mean metered production load at that hour over the $N$ preceding similar days, that is, working days or non-working days. Since the PV output offsets the production load behind the meter, the metered load is the production load minus the PV output:
\begin{equation}
L^{\mathrm{base}}(h) = \frac{1}{N} \sum_{i=1}^{N} \bigl(L_{i}^{\mathrm{pro}}(h) - P_{i}^{\mathrm{PV}}(h)\bigr),
\label{eq:baseline}
\end{equation}
where $L_{i}^{\mathrm{pro}}(h)$ and $P_{i}^{\mathrm{PV}}(h)$ are the production load and the PV output at hour $h$ on the $i$-th preceding similar day. We write $L^{\mathrm{base}}(t)$ for the baseline at the hour of time slot $t$. The PV energy and the battery energy allocated to the production zone offset the production load, and the response energy $\Delta L(t)$ is the resulting reduction below the baseline:
\begin{equation}
\begin{split}
\Delta L(t) &= L^{\mathrm{base}}(t)\Delta t - \bigl[\bigl(L^{\mathrm{pro}}(t) - P^{\mathrm{PV}}(t)\bigr)\Delta t\\
&\quad - r_{\mathrm{dis}}(t)\, Q^{\mathrm{bat}}(t)\bigr]^{+},
\end{split}
\label{eq:adjusted_load}
\end{equation}
where $[\cdot]^{+} = \max(0,\cdot)$, $Q^{\mathrm{bat}}(t) = -\bigl(P^{\mathrm{ESS}}(t) + P^{\mathrm{EV}}(t)\bigr)\Delta t$ is the net battery discharge energy, and the distribution ratio $r_{\mathrm{dis}}(t) \in [0,1]$ allocates it to the production zone. The office zone receives the remaining share. The best allocation depends on whether the batteries charge or discharge and on whether the production load is already offset, so $r_{\mathrm{dis}}(t)$ is a decision variable rather than a fixed share. The response ratio $r_{\mathrm{dr}}(t) = \Delta L(t)/L_{\mathrm{dr}}$ relates the response energy to the invited response energy per time slot $L_{\mathrm{dr}}$, and the revenue follows the tiered compensation of the plan:
\begin{equation}
R_{\mathrm{dr}}(t) = d(t)
\begin{cases}
0, & r_{\mathrm{dr}}(t) < \beta_1, \\
\gamma\, p_{\mathrm{dr}}\, \Delta L(t), & \beta_1 \le r_{\mathrm{dr}}(t) \le \beta_2, \\
p_{\mathrm{dr}}\, \Delta L(t), & \beta_2 < r_{\mathrm{dr}}(t) \le \beta_3, \\
\beta_3\, p_{\mathrm{dr}}\, L_{\mathrm{dr}}, & r_{\mathrm{dr}}(t) > \beta_3,
\end{cases}
\label{eq:dr_revenue}
\end{equation}
where $d(t) \in \{0,1\}$ indicates a DR event, $p_{\mathrm{dr}}$ is the unit compensation price, $\beta_1 < \beta_2 < \beta_3$ are the response thresholds and $\gamma < 1$ is the compensation coefficient of a partial response. A response above $\beta_3 L_{\mathrm{dr}}$ is thus compensated as $\beta_3 L_{\mathrm{dr}}$.

\subsubsection{Grid and Carbon Costs}
Imports are charged and exports credited at the time-of-use (TOU) price $p(t)$. The carbon cost is charged to imported electricity:
\begin{equation}
\begin{split}
C_{\mathrm{grid}}(t) &= p(t)\, P^{\mathrm{grid}}(t)\, \Delta t,\\
C_{\mathrm{CO_2}}(t) &= p_{\mathrm{c}}\, \omega\, \bigl[P^{\mathrm{grid}}(t)\Delta t\bigr]^{+},
\end{split}
\label{eq:grid_carbon}
\end{equation}
where $p_{\mathrm{c}}$ is the carbon price and $\omega$ is the carbon intensity of grid electricity.

\subsubsection{Battery Ageing Cost}
The 2 chemistries age through different mechanisms, so separate empirical cycle-ageing models are adopted. The relative capacity loss of the LFP cells is modelled as follows~\cite{naumann2020analysis}:
\begin{equation}
\begin{split}
Q^{\mathrm{LFP}}_{\mathrm{loss}} &= 10^{-2}\bigl(a_1 C_{\mathrm{rate}} + a_2\bigr)\\
&\quad \times \bigl(a_3(\mathrm{DOD} - 0.6)^3 + a_4\bigr) N_{\mathrm{EFC}}^{\,z},
\end{split}
\label{eq:lfp}
\end{equation}
where $C_{\mathrm{rate}}$ is the C-rate, DOD is the depth of discharge, $N_{\mathrm{EFC}}$ is the number of equivalent full cycles (EFCs), $a_1$ to $a_4$ and $z$ are fitted parameters, and the factor $10^{-2}$ converts the fitted loss from percent to a fraction. The relative capacity loss of the NMC cells is modelled as follows~\cite{schmalstieg2014holistic}:
\begin{equation}
Q^{\mathrm{NMC}}_{\mathrm{loss}} = \bigl(b_1(V_{\mathrm{avg}} - V_0)^2 + b_2 + b_3\,\mathrm{DOD}\bigr)\sqrt{N_{\mathrm{EFC}}\, C_0},
\label{eq:nmc}
\end{equation}
where $V_{\mathrm{avg}}$ is the average cell voltage, $V_0$ is the reference voltage, $C_0$ is the nominal cell capacity in ampere-hours and $b_1$ to $b_3$ are fitted parameters. The fitted parameters of both models are listed in Table~\ref{tab:parameters_system}. To price ageing at the dispatch level, 1 EFC is taken to move an energy throughput of $2E^j$ at a reference battery age of $N_{\mathrm{ref}}$ EFCs, which gives the ageing cost per kWh of throughput and the ageing cost of each time slot:
\begin{equation}
\alpha^j = \frac{\Delta Q^j\, C^j_{\mathrm{kWh}}}{2}, \quad C_{\mathrm{deg}}(t) = \sum_{j} \alpha^j \bigl|P^j(t)\bigr| \Delta t,
\label{eq:alpha}
\end{equation}
where $\Delta Q^j = Q^j_{\mathrm{loss}}(N_{\mathrm{ref}}+1) - Q^j_{\mathrm{loss}}(N_{\mathrm{ref}})$ is the marginal relative capacity loss, evaluated at the rated power and over the full SoC window, and $C^j_{\mathrm{kWh}}$ is the battery price per kWh of capacity.

\subsubsection{Constraint Violations}
The comfort constraint in~\eqref{eq:comfort} and the departure constraint~\eqref{eq:departure} are handled as constraints of the learning problem, and their violations are measured as follows:
\begin{equation}
\begin{gathered}
C_{\mathrm{T}}(t) = [T_{\mathrm{in}}(t+1)-T_{\max}]^{+} + [T_{\min}-T_{\mathrm{in}}(t+1)]^{+}\\
C_{\mathrm{SoC}}(t) = \bigl[\text{SoC}_{\mathrm{lim}} - \text{SoC}^{\mathrm{EV}}(t+1)\bigr]^{+}\, \mathbf{1}\{t \in \mathcal{T}_{\mathrm{dep}}\},
\end{gathered}
\label{eq:violation}
\end{equation}
where $\mathbf{1}\{\cdot\}$ is the indicator function.

\section{Proposed Method}\label{sec:Proposed method}

\begin{figure*}[tb]
\centering
\includegraphics[width=0.96\textwidth]{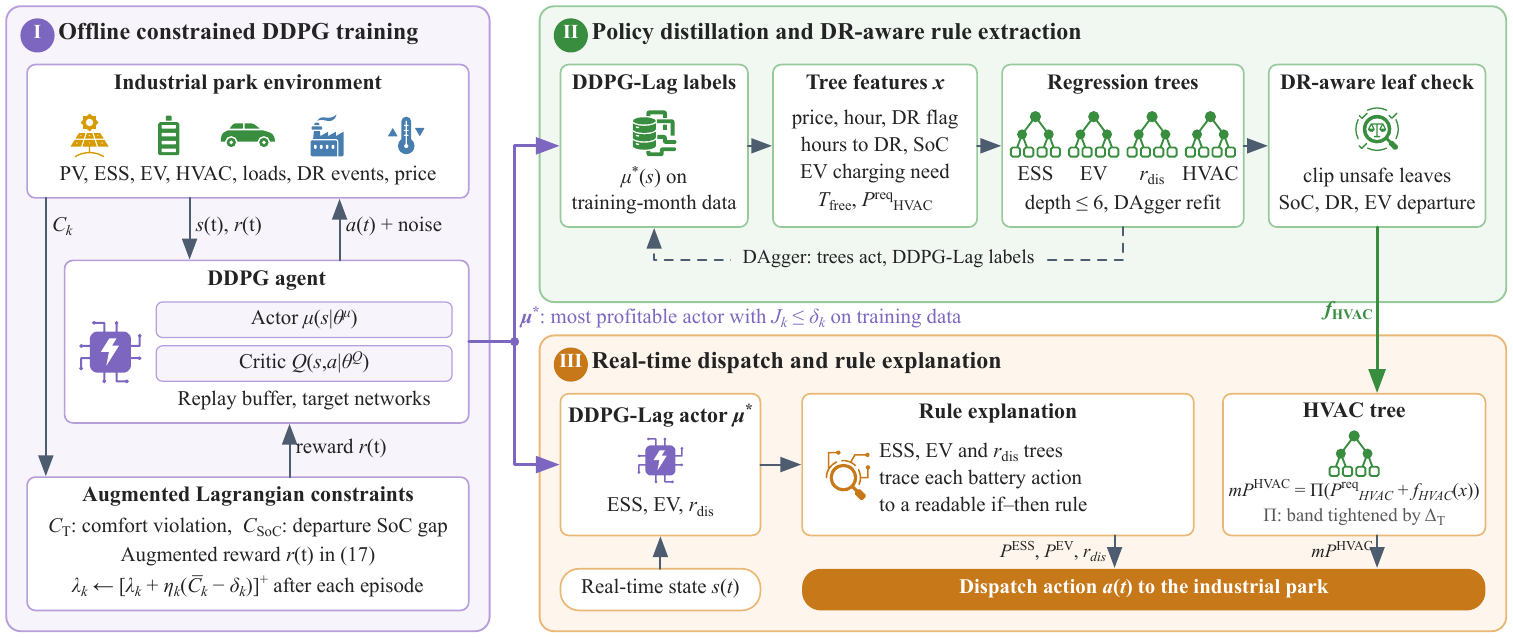}
\caption{Framework of the proposed DDPG-Lag-DT method.}
\label{fig:algorithm framework}
\end{figure*}

The proposed framework is shown in Fig.~\ref{fig:algorithm framework}. It consists of 3 stages. First, DDPG-Lag learns the ESS power, EV power, distribution ratio and HVAC power offline. Second, the complete policy is distilled into 4 regression trees with physics-informed features. Finally, at deployment, DDPG-Lag dispatches the batteries and sets the distribution ratio, the residual HVAC tree controls office temperature, and the battery trees summarise the learned dispatch.

\subsection{Constrained Learning and DDPG-Lag}

The dispatch problem~\eqref{eq:problem} is formulated as a constrained Markov decision process (CMDP)~\cite{altman1999constrained}. The state collects the information available to the EMS at time slot $t$:
\begin{equation}
\begin{split}
s(t) &= \bigl\{ p(t), L^{\mathrm{pro}}(t), L^{\mathrm{off}}(t), P^{\mathrm{PV}}(t), L^{\mathrm{base}}(t),\\
& \hat{T}_{\mathrm{out}}(t), T_{\mathrm{in}}(t), \text{SoC}^{\mathrm{ESS}}(t), \text{SoC}^{\mathrm{EV}}(t),\\
& \sin\tfrac{2\pi\varphi(t)}{H}, \cos\tfrac{2\pi\varphi(t)}{H}, d(t), e(t), h_{\mathrm{dep}}(t), w(t) \bigr\},
\end{split}
\label{eq:state}
\end{equation}
where $\hat{T}_{\mathrm{out}}(t)$ is the forecast of the outdoor temperature, $\varphi(t)$ is the hour of the day, $h_{\mathrm{dep}}(t)$ is the number of hours until the scheduled departure of the EVs and $w(t)$ indicates a working day. The announced DR schedule, measured EV plug status and scheduled departure provide the available timing information. The actor produces 4 normalised commands in $[-1,1]$ for the ESS power, EV power, distribution ratio and signed HVAC power. The charging-positive battery commands are $\tilde{P}^{j}(t)=-a_j(t)P^j_{\mathrm{rated}}$, where $a_j(t)$ is the corresponding actor output. The ratio is mapped as $r_{\mathrm{dis}}(t)=(a_r(t)+1)/2$, and the signed HVAC command is $a_{\mathrm H}(t)P^{\mathrm{HVAC}}_{\max}$, where $a_r(t)$ and $a_{\mathrm H}(t)$ are the remaining actor outputs. The battery powers are clipped to the limits in~\eqref{eq:battery_limits}, and the grid power follows from~\eqref{powerblance}.

The reward is the net operating profit minus an augmented Lagrangian penalty on each violation:
\begin{equation}
\begin{split}
r(t) = \frac{1}{\sigma}\Bigl[&R_{\mathrm{dr}}(t) - C_{\mathrm{grid}}(t) - C_{\mathrm{CO_2}}(t) - C_{\mathrm{deg}}(t)\\
&- \sum_{k \in \{\mathrm{T},\mathrm{SoC}\}} \Bigl(\lambda_k C_k(t) + \frac{\rho_k}{2} C_k(t)^2\Bigr)\Bigr],
\end{split}
\label{eq:aug_reward}
\end{equation}
where $\lambda_k \ge 0$ are the Lagrange multipliers of the 2 constraints, $\rho_k > 0$ are the quadratic penalty coefficients and $\sigma$ scales the reward. Let $J_{\mathrm{E}}(\mu)$ denote the expected return of the economic terms under the policy $\mu$, and let $J_k(\mu)$ denote the expected mean violation $C_k$ per episode. The CMDP is formulated as follows:
\begin{equation}
\max_{\mu}\ J_{\mathrm{E}}(\mu) \quad \text{s.t.}\ J_k(\mu) \le \delta_k, \quad k \in \{\mathrm{T},\mathrm{SoC}\},
\label{eq:cmdp}
\end{equation}
where $\delta_k$ is the constraint tolerance. Lagrangian methods alternate between policy optimisation and multiplier updates~\cite{tessler2019reward}. The linear term of~\eqref{eq:aug_reward} adaptively weights each violation, and the quadratic term penalises large violations. Policy optimisation is performed by DDPG with a deterministic actor and an action-value critic~\cite{lillicrap2016continuous}. Ornstein--Uhlenbeck (OU) noise supports exploration. After each episode, the multipliers follow the projected dual-ascent step:
\begin{equation}
\lambda_k \leftarrow \bigl[\lambda_k + \eta_k\bigl(\bar{C}_k - \delta_k\bigr)\bigr]^{+}, \quad k \in \{\mathrm{T},\mathrm{SoC}\},
\label{eq:dual}
\end{equation}
where $\eta_k$ is the dual step size and $\bar{C}_k$ is the episode mean violation, averaged over hours for comfort and departures for EVs. A multiplier increases when its violation exceeds the tolerance and decreases when the constraint holds with a margin. Final policy selection retains the highest-profit actor that satisfies the tolerances on the training data. The selected actor is denoted by $\mu^\ast$, and the resulting agent is termed DDPG-Lag.

\subsection{Physics-Informed Policy Distillation}

\begin{figure}[tb]
\centering
\includegraphics[width=0.95\columnwidth]{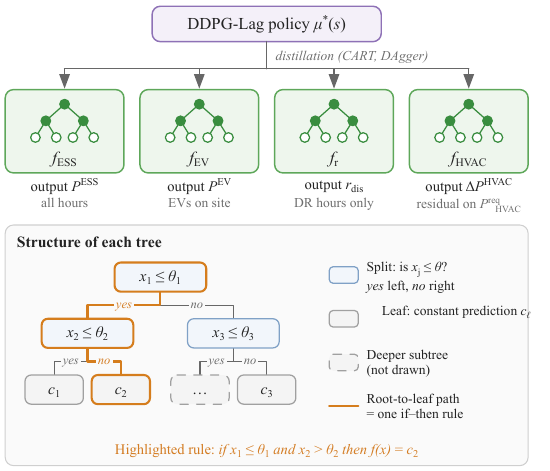}
\caption{Policy distillation and decision-tree structure.}
\label{fig:dt_schematic}
\end{figure}

The DDPG-Lag policy $\mu^\ast$ is distilled into 4 regression trees, $f_{\mathrm{ESS}}$, $f_{\mathrm{EV}}$, $f_r$ and $f_{\mathrm{HVAC}}$, as illustrated in Fig.~\ref{fig:dt_schematic}. The classification and regression tree (CART) algorithm fits normalised action and HVAC residual targets by minimising squared prediction error, with cost-complexity pruning coefficient $\alpha_{\mathrm{cc}}$~\cite{breiman1984classification}. Predictions are mapped to physical ranges. Each root-to-leaf path gives an if--then rule with a constant prediction.

The 4 model-derived features represent the physical scheduling requirements. First, $h_{\mathrm{DR}}(t)$ encodes the time to the next announced DR hour of the same day, with zero identifying an ongoing event. Second, the required EV charging rate is the mean charging power, as a fraction of rated power, that reaches the departure target:
\begin{equation}
P^{\mathrm{req}}_{\mathrm{EV}}(t) = \frac{e(t)\bigl[\text{SoC}_{\mathrm{lim}} - \text{SoC}^{\mathrm{EV}}(t)\bigr]^{+} E^{\mathrm{EV}}}{\eta^{\mathrm{EV}}_{\mathrm{c}} \max\{h_{\mathrm{dep}}(t), \Delta t\}\, P^{\mathrm{EV}}_{\mathrm{rated}}}.
\label{eq:preq_ev}
\end{equation}
Third, the free-floating temperature $T_{\mathrm{free}}(t)=\varepsilon T_{\mathrm{in}}(t)+(1-\varepsilon)\hat T_{\mathrm{out}}(t)$ is the predicted indoor temperature of the next slot with the HVAC off. Finally, the signed HVAC powers that reach the comfort bounds tightened by margin $\Delta_{\mathrm T}$ against forecast errors follow from~\eqref{eq:temperature}:
\begin{equation}
\begin{split}
P^{\mathrm{lo}}_{\mathrm{HVAC}}(t) &= \frac{G_{\mathrm{b}}\bigl(T_{\min} + \Delta_{\mathrm{T}} - T_{\mathrm{free}}(t)\bigr)}{\eta^{\mathrm{HVAC}}(1-\varepsilon)},\\
P^{\mathrm{hi}}_{\mathrm{HVAC}}(t) &= \frac{G_{\mathrm{b}}\bigl(T_{\max} - \Delta_{\mathrm{T}} - T_{\mathrm{free}}(t)\bigr)}{\eta^{\mathrm{HVAC}}(1-\varepsilon)}.
\end{split}
\label{eq:preq_hvac}
\end{equation}
Let $\Pi(x)=\min\{\max\{x,P^{\mathrm{lo}}_{\mathrm{HVAC}}(t)\},P^{\mathrm{hi}}_{\mathrm{HVAC}}(t)\}$ project onto this interval. The required-power feature $P^{\mathrm{req}}_{\mathrm{HVAC}}(t)$ is $\Pi(0)$ bounded by $\pm\kappa_{\mathrm f}P^{\mathrm{HVAC}}_{\max}$, where $\kappa_{\mathrm f}>1$. It is zero inside the tightened band and positive for heating. These features are supplied to the trees but not to DDPG-Lag. The HVAC tree predicts $\hat P^{\mathrm{HVAC}}(t)=P^{\mathrm{req}}_{\mathrm{HVAC}}(t)+f_{\mathrm{HVAC}}(x_{\mathrm{HVAC}}(t))$, so each leaf specifies a residual adjustment, with $x_{\mathrm{HVAC}}(t)$ denoting its feature vector. The allocation tree is fitted on DR hours because $r_{\mathrm{dis}}$ affects DR revenue through~\eqref{eq:adjusted_load} rather than grid power. The EV tree uses on-site states.

The trees are fitted to trajectories labelled by DDPG-Lag and refined through dataset aggregation (DAgger)~\cite{ross2011reduction}. During aggregation, DDPG-Lag labels the states visited by the trees, extending coverage of their operating trajectories. Battery leaf predictions are checked against SoC, DR and departure operating rules. The offline procedure is summarised in Algorithm~\ref{alg:offline}.

\begin{algorithm}[tb]
\caption{Offline training and distillation of DDPG-Lag}
\label{alg:offline}
\begin{algorithmic}[1]
\REQUIRE Training data, $\lambda^0_k$, $N_{\mathrm{ep}}$, $N_{\mathrm{eval}}$, $N_{\mathrm{T}}$, $N_{\mathrm{D}}$, $N_{\mathrm{P}}$
\STATE Initialise the actor $\mu$, the critic, their target networks, the replay buffer and $\lambda_k \leftarrow \lambda^0_k$
\FOR{episode $= 1$ to $N_{\mathrm{ep}}$}
\STATE Run $\mu$ with OU noise over a random episode of the training data, store the transitions with the reward~\eqref{eq:aug_reward}, and update the critic, the actor and the target networks at every step
\STATE Update $\lambda_{\mathrm{T}}$ and $\lambda_{\mathrm{SoC}}$ by~\eqref{eq:dual}
\STATE Every $N_{\mathrm{eval}}$ episodes, evaluate $\mu$ on the training data and update $\mu^{\ast}$ by constraint-priority policy selection
\ENDFOR
\STATE Run $\mu^{\ast}$ with Gaussian action noise for $N_{\mathrm{T}}$ passes, and store the tree features and DDPG-Lag actions of the visited states in $\mathcal{B}$
\FOR{$i = 1$ to $N_{\mathrm{D}}$}
\STATE Fit the 4 trees on $\mathcal{B}$, apply the leaf check, run the trees for $N_{\mathrm{P}}$ passes and add the visited states labelled by $\mu^{\ast}$ to $\mathcal{B}$
\ENDFOR
\STATE Fit the final trees on $\mathcal{B}$ and apply the leaf check
\RETURN $\mu^{\ast}$, $f_{\mathrm{ESS}}$, $f_{\mathrm{EV}}$, $f_{r}$ and $f_{\mathrm{HVAC}}$
\end{algorithmic}
\end{algorithm}

\subsection{Deployment}

At deployment, DDPG-Lag determines the ESS power, EV power and distribution ratio. The residual HVAC command is projected onto the model-predicted power interval that maintains the tightened comfort band:
\begin{equation}
m(t)P^{\mathrm{HVAC}}(t) = \Pi\bigl(P^{\mathrm{req}}_{\mathrm{HVAC}}(t) + f_{\mathrm{HVAC}}(x_{\mathrm{HVAC}}(t))\bigr),
\label{eq:hvac_tree}
\end{equation}
which is bounded by the rated power in~\eqref{eq:comfort}. The resulting DDPG-Lag-DT controller combines learned battery dispatch with executable HVAC rules. The ESS, EV and allocation trees provide readable summaries of the battery policy.

The hybrid controller separates one-step comfort control from intertemporal battery scheduling. The thermal model gives the admissible HVAC powers in closed form. DDPG-Lag constrains episode-mean violations, while the projection enforces the model-predicted tightened band at each slot whenever rated power permits. Each HVAC command is computed from a tree path and the comfort projection. The DDPG-Lag critic accounts for future prices, DR events and departure requirements when scheduling the batteries. Deployment is summarised in Algorithm~\ref{alg:deploy}.

\begin{algorithm}[tb]
\caption{DDPG-Lag-DT energy management}
\label{alg:deploy}
\begin{algorithmic}[1]
\REQUIRE Actor $\mu^{\ast}$, trees $f_{\mathrm{ESS}}$, $f_{\mathrm{EV}}$, $f_{r}$ and $f_{\mathrm{HVAC}}$
\FOR{each time slot $t$}
\STATE Observe $s(t)$, and compute $T_{\mathrm{free}}(t)$, the HVAC power bounds~\eqref{eq:preq_hvac}, $P^{\mathrm{req}}_{\mathrm{HVAC}}(t)$ and the tree features $x_{\mathrm{HVAC}}(t)$
\STATE Obtain $P^{\mathrm{ESS}}(t)$, $P^{\mathrm{EV}}(t)$ and $r_{\mathrm{dis}}(t)$ from $\mu^{\ast}(s(t))$
\STATE Set the HVAC power by~\eqref{eq:hvac_tree}
\STATE Apply the dispatch, with the powers clipped to the limits in~\eqref{eq:comfort} and~\eqref{eq:battery_limits}
\STATE Summarise the battery dispatch by the rules of $f_{\mathrm{ESS}}$, $f_{\mathrm{EV}}$ and $f_{r}$ that the current features satisfy
\ENDFOR
\end{algorithmic}
\end{algorithm}

\section{Case Studies}\label{sec:Results}

\subsection{Case Study Setup}

The load and PV data were collected from an EMS-equipped industrial park in Fujian Province, China, which manufactures solar cables. Its production load peaks at about 266~kW, and its office base load is 20~kW. The case-study TOU price profile has a valley price of about 0.40 Chinese yuan (RMB) per kWh from 00:00 to 08:00 and peak and sharp prices of 0.90 to 1.03~RMB/kWh at 10:00 to 12:00, 15:00 to 20:00 and 21:00 to 22:00. The agents are trained on 168-h episodes drawn from the 672~h of 1 to 28 September 2024 and evaluated on the 720~h of September 2025.

The DR compensation tiers follow the Fujian electric power DR implementation plan issued in July 2024~\cite{fujian2024demandresponse}. DR events take place on working days from 10:00 to 13:00, on 20 days of the training month and 22 days of the test month. The plan's day-ahead response-speed coefficient is 1. The case-study net baseline subtracts the mean PV output of up to 5 preceding working days or 3 preceding non-working days from the policy-screened production-load baseline. On the first working and non-working day of each month, the PV reference uses the following comparable days. All methods share this precomputed baseline. The tiers in~\eqref{eq:dr_revenue} are $\beta_1 = 0.5$, $\beta_2 = 0.8$, $\beta_3 = 2$ and $\gamma = 0.6$, so a response above 200\% of the invited load is counted as 200\%. The plan leaves the unit price to an annual auction, so $p_{\mathrm{dr}}$ is set in this study to 3 times the TOU price of the DR hour, which gives 1.97 to 2.96~RMB/kWh in the test month.

The EVs arrive at 08:00 and are scheduled to leave at 19:00 every day, which gives 30 departures in the test month. The EVs arrive at $\text{SoC}_{\mathrm{arr}}$ each day. To reflect the information available at dispatch, the EMS observes an outdoor temperature forecast with a Gaussian hourly error of standard deviation 1\,$^\circ$C, and the EVs leave 0 or 1~h before the scheduled time with equal probability each day. New realisations are drawn in every training episode, and all methods are evaluated on the same 5 realisations of the test month. The case study runs at a time step of $\Delta t = 1$~h.

The battery prices follow the 2025 BloombergNEF survey, in which stationary storage packs cost 70 US dollars (USD) per kWh and NMC packs cost 128~USD/kWh~\cite{bnef2025battery}. An exchange rate of 7.18~RMB/USD gives the rounded battery prices and, by~\eqref{eq:alpha}, the ageing cost coefficients in Table~\ref{tab:parameters_system}, which lists the main parameters. Twin delayed DDPG (TD3)~\cite{fujimoto2018addressing} and soft actor--critic (SAC)~\cite{haarnoja2018soft} are used as learning comparators. All DRL actors and critics have 2 hidden layers of 256 rectified linear units, with actor and critic learning rates of $10^{-4}$ and $10^{-3}$ and a discount factor of 0.99. Distillation uses training trajectories, with the last 20\% of DDPG-Lag passes reserved for fidelity evaluation. Each evaluation starts from $\text{SoC}^{\mathrm{ESS}} = 0.5$ and $T_{\mathrm{in}} = 22\,^\circ$C.

\begin{table}[tb]
\centering
\caption{Main parameters}
\label{tab:parameters_system}
\setlength{\tabcolsep}{2.9pt}
\renewcommand{\arraystretch}{1.15}
\begin{tabular}{llc}
\toprule
Component & Parameter & Value \\
\midrule
\multirow{3}{*}{HVAC} & $T_{\min}$, $T_{\max}$, $\Delta_{\mathrm{T}}$ & 19, 24, 0.5\,$^\circ$C \\
 & $\varepsilon$, $\eta^{\mathrm{HVAC}}$, $G_{\mathrm{b}}$ & 0.85, 3.2, 18 kW/$^\circ$C \\
 & $P^{\mathrm{HVAC}}_{\max}$ & 50 kW \\
\addlinespace
\multirow{3}{*}{ESS, EV} & $E^j$, $P^j_{\mathrm{rated}}$, $P^j_{\mathrm{sb}}$ & 400 kWh, 100 kW, 0.99 kW \\
 & $\text{SoC}^j_{\min}$, $\text{SoC}^j_{\max}$ & 0.2, 1.0 \\
 & $\eta^j_{\mathrm{c}}$, $\eta^j_{\mathrm{d}}$, $N_{\mathrm{ref}}$ & 0.95, 0.95, 365 EFCs \\
\addlinespace
\multirow{3}{*}{ESS (LFP)} & $C^{\mathrm{ESS}}_{\mathrm{kWh}}$, $\alpha^{\mathrm{ESS}}$ & 500, 0.0083 RMB/kWh \\
 & $a_1$, $a_2$ & 0.0630, 0.0971 \\
 & $a_3$, $a_4$, $z$ & 4.0253, 1.0923, 0.5 \\
\addlinespace
\multirow{4}{*}{EV (NMC)} & $C^{\mathrm{EV}}_{\mathrm{kWh}}$, $\alpha^{\mathrm{EV}}$ & 920, 0.0695 RMB/kWh \\
 & $\text{SoC}_{\mathrm{arr}}$, $\text{SoC}_{\mathrm{lim}}$ & 0.35, 0.6 \\
 & $b_1$, $b_2$, $b_3$ ($\times10^{-3}$) & 7.348, 0.760, 4.081 \\
 & $V_0$, $V_{\mathrm{avg}}$, $C_0$ & 3.667 V, 3.7 V, 2.05 Ah \\
\addlinespace
\multirow{2}{*}{DR, carbon} & $L_{\mathrm{dr}}$, $p_{\mathrm{dr}}$ & 100 kWh, $3p(t)$ \\
 & $\omega$, $p_{\mathrm{c}}$ & 0.28088 kg/kWh, 0.06 RMB/kg \\
\addlinespace
\multirow{4}{*}{DDPG-Lag} & $N_{\mathrm{ep}}$, $N_{\mathrm{eval}}$ & 7500, 100 \\
 & $\sigma$, $\lambda^0_{\mathrm{T}}$, $\lambda^0_{\mathrm{SoC}}$ & 100, 48, 510 \\
 & $\rho_{\mathrm{T}}$, $\rho_{\mathrm{SoC}}$, $\eta_{\mathrm{T}}$, $\eta_{\mathrm{SoC}}$ & 1000, 3000, 5, 100 \\
 & $\delta_{\mathrm{T}}$, $\delta_{\mathrm{SoC}}$ & 0.01, 0.01 \\
\addlinespace
\multirow{7}{*}{Trees} & Depth, leaf size, $\alpha_{\mathrm{cc}}$ & 6, 50, 0.002 \\
 & $N_{\mathrm{T}}$, $N_{\mathrm{D}}$, $N_{\mathrm{P}}$ & 30, 3, 10 \\
 & Action noise, holdout & 0.1, 20\% \\
 & Minimum impurity decrease & 10$^{-4}$ \\
 & HVAC feature limit $\kappa_{\mathrm f}$ & 1.5 \\
 & EV departure-rule window & 3 h \\
 & Leaf check: SoC, power & 0.9, 0.25, 30 kW \\
\addlinespace
Uncertainty & Forecast error, early leave & 1\,$^\circ$C, 0--1 h \\
\bottomrule
\end{tabular}
\end{table}

The proposed method, DDPG-Lag-DT, is compared with rule-based control, SAC, TD3 and DDPG. The rule-based strategy discharges the ESS at full power during DR events and charges it at the valley price. It charges the EVs at full power whenever they are on site, allocates all battery energy to the production zone during DR events, and keeps $T_{\mathrm{in}}(t+1)$ at least 0.5\,$^\circ$C inside the comfort band with a predictive thermostat, the same margin as $\Delta_{\mathrm{T}}$. SAC, TD3 and DDPG use the same state, action, reward and policy selection as DDPG-Lag, but their multipliers stay at the initial values, which corresponds to fixed penalty weights. All DRL methods are trained with 5 random seeds, and each seed is evaluated on the 5 realisations. Evaluation uses 4 test-month metrics: the net operating cost, the number of hours in which $T_{\mathrm{in}}(t+1)$ leaves the comfort band, the number of the 30 EV departures below $\text{SoC}_{\mathrm{lim}}$, and the DR response rate, which is the share of DR hours with $r_{\mathrm{dr}}(t) \ge 0.5$ that earn compensation. Differences between methods are tested with Welch's $t$-test over the seeds.

\subsection{Training Convergence}

\begin{figure}[tb]
\centering
\includegraphics[width=\columnwidth]{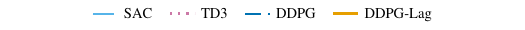}\\[-4pt]
\subfloat[]{\includegraphics[width=0.49\columnwidth]{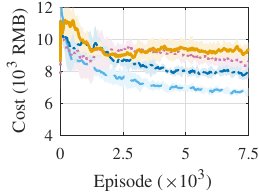}\label{fig:train_cost}}
\hfill
\subfloat[]{\includegraphics[width=0.49\columnwidth]{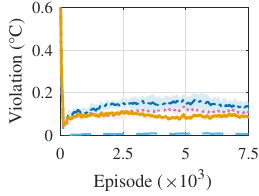}\label{fig:train_temp}}
\\
\subfloat[]{\includegraphics[width=0.49\columnwidth]{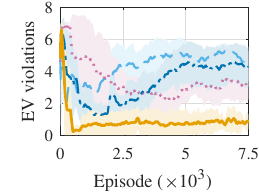}\label{fig:train_ev}}
\hfill
\subfloat[]{\includegraphics[width=0.49\columnwidth]{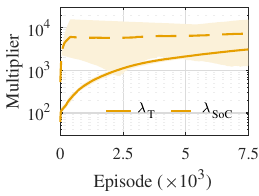}\label{fig:train_lambda}}
\caption{Training convergence: (a) cost, (b) temperature violations, (c) EV departure violations, (d) multipliers.}
\label{fig:rewards_DDPG}
\end{figure}

Constraint-driven learning reduces EV departure shortfalls during training, as shown in Fig.~\ref{fig:rewards_DDPG}. The 100-episode moving averages are presented as 5-seed means with bands of 1 standard deviation in Fig.~\ref{fig:rewards_DDPG}(a)--(c). The vertical axis in Fig.~\ref{fig:rewards_DDPG}(b) is restricted to 0--0.6\,$^\circ$C. DDPG-Lag reduces departure violations to below 1 per week within the first 500 episodes and holds them at about 0.8 per week, whereas SAC, TD3 and DDPG settle at 3.2 to 4.7. Its mean temperature violation is 0.09\,$^\circ$C, below the 0.11 and 0.14\,$^\circ$C of TD3 and DDPG. The mean multipliers and their seed ranges are shown in Fig.~\ref{fig:rewards_DDPG}(d). Dual ascent raises $\lambda_{\mathrm{T}}$ from 48 to about 3100 and $\lambda_{\mathrm{SoC}}$ from 510 to between 1300 and 15\,800, bringing both penalties to the operating-cost scale.

\subsection{Performance Comparison}

\begin{table*}[tb]
\centering
\caption{Performance on the test month and ablation of DDPG-Lag}
\label{tab:performance_comparison}
\setlength{\tabcolsep}{10pt}
\begin{tabular}{lcccc}
\toprule
Method & \shortstack{Net operating\\cost (RMB)} & \shortstack{Temperature\\violations (h)} & \shortstack{EV departure\\violations (of 30)} & \shortstack{DR response\\rate (\%)} \\
\midrule
Rule-based & 39\,166 & 0.2 & 0 & 66.7 \\
SAC & 31\,104\,$\pm$\,852 & 77.5\,$\pm$\,32.0 & 13.2\,$\pm$\,7.4 & 76.2\,$\pm$\,7.7 \\
TD3 & 38\,420\,$\pm$\,5617 & 39.9\,$\pm$\,30.4 & 2.6\,$\pm$\,3.4 & 70.8\,$\pm$\,23.1 \\
DDPG & 35\,986\,$\pm$\,819 & 44.4\,$\pm$\,12.2 & 2.2\,$\pm$\,0.8 & 83.3\,$\pm$\,2.6 \\
DDPG-Lag & 36\,410\,$\pm$\,3298 & 36.8\,$\pm$\,38.9 & 0.6\,$\pm$\,0.8 & 77.9\,$\pm$\,5.6 \\
\textbf{Proposed method} & \textbf{35\,429}\,$\pm$\,\textbf{2211} & \textbf{0.04}\,$\pm$\,\textbf{0.09} & \textbf{0.6}\,$\pm$\,\textbf{0.8} & \textbf{77.4}\,$\pm$\,\textbf{6.2} \\
\midrule
\multicolumn{5}{l}{Ablation of DDPG-Lag} \\
Small fixed step & 41\,671\,$\pm$\,1816 & 27.9\,$\pm$\,31.1 & 3.2\,$\pm$\,4.1 & 67.1\,$\pm$\,6.4 \\
Small adaptive step & 36\,270\,$\pm$\,3289 & 59.3\,$\pm$\,23.8 & 2.0\,$\pm$\,2.4 & 77.6\,$\pm$\,6.3 \\
Small adaptive step, $r_{\mathrm{dis}} = 0.5$ & 38\,828\,$\pm$\,4108 & 61.0\,$\pm$\,25.8 & 4.0\,$\pm$\,2.8 & 70.9\,$\pm$\,15.9 \\
Small adaptive step, no ageing cost & 40\,687\,$\pm$\,9135 & 48.5\,$\pm$\,26.9 & 1.3\,$\pm$\,2.3 & 63.1\,$\pm$\,23.5 \\
\bottomrule
\end{tabular}
\end{table*}

The proposed method combines constraint satisfaction with lower operating cost, as summarised in Table~\ref{tab:performance_comparison}. Averaged over the 5 seeds and 5 realisations, comfort violations are 0.04~h and EV departure compliance is 98\%, corresponding to 0.6 violations out of 30. Its net operating cost of 35\,429~RMB is 9.5\% below that of the rule-based strategy, whose thermostat uses the same comfort margin. Battery scheduling increases DR revenue by 28.8\% and reduces the electricity bill by 3.6\% through DR participation and evening-peak discharge.

Relative to fixed-penalty SAC, TD3 and DDPG, the proposed method reduces comfort violations by 99.9\%, with p-values at most 0.043, and departure violations by 72.7\% to 95.4\%. Its cost is 7.8\% and 1.5\% below TD3 and DDPG. SAC's lower cost of 31\,104~RMB is accompanied by 77.5~h of comfort violations and 13.2 departure shortfalls.

Constraint adaptation improves EV departure compliance, while residual HVAC deployment improves thermal comfort. DDPG-Lag reduces EV departure violations by 72.7\% relative to DDPG, with a p-value of 0.015. Battery dispatch is determined by DDPG-Lag, while the residual HVAC tree supplies the required cooling plus a learned margin. It reduces comfort violations from 36.8 to 0.04~h, improving every seed, and lowers cost by 2.7\%. On the first test realisation, it uses 12.4\% less HVAC energy than DDPG-Lag.

\subsection{Scheduling Behaviour}

\begin{figure}[tb]
\centering
\subfloat[]{\includegraphics[width=\columnwidth]{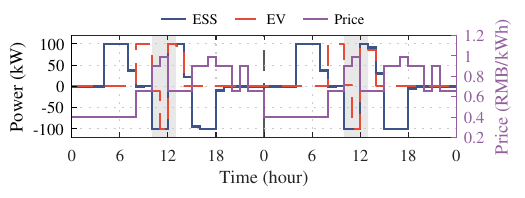}\label{fig:state_sub_a}}
\\
\subfloat[]{\includegraphics[width=\columnwidth]{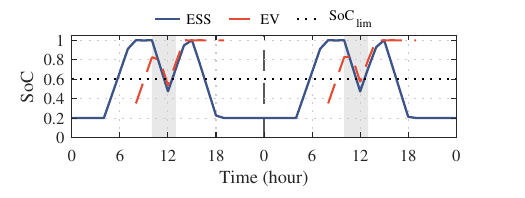}\label{fig:state_sub_b}}
\\
\subfloat[]{\includegraphics[width=\columnwidth]{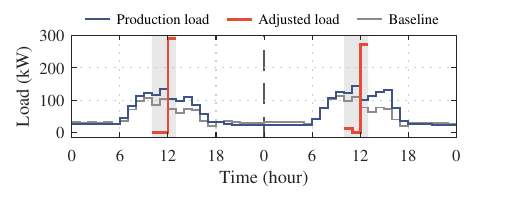}\label{fig:state_sub_c}}
\caption{Dispatch: (a) battery power and price, (b) SoC, (c) production load, adjusted load and DR baseline.}
\label{fig:state}
\end{figure}

\begin{figure}[tb]
\centering
\subfloat[]{\includegraphics[width=\columnwidth]{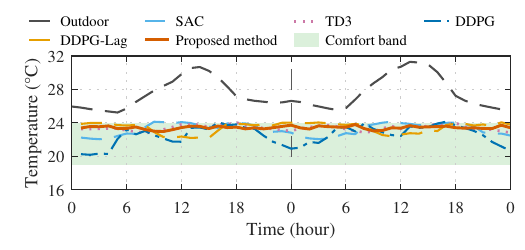}\label{fig:temp_control}}
\\
\subfloat[]{\includegraphics[width=\columnwidth]{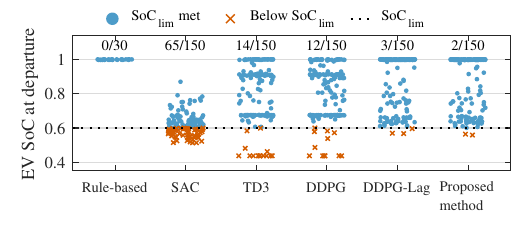}\label{fig:departure_soc}}
\caption{Constraint satisfaction: (a) indoor temperature, (b) EV departure SoC.}
\label{fig:soc_ev}
\end{figure}

The learned dispatch combines DR participation with evening-peak arbitrage, as shown in Fig.~\ref{fig:state} for 9 and 10 September 2025, the first seed and realisation. Positive battery power denotes charging, and grey shading marks DR. The ESS charges at the valley price until it is full at 08:00, and EVs arrive at an SoC of 0.35. During the first 2 DR hours, both batteries discharge at up to 100~kW and allocate this energy to production, reducing its adjusted load to 0--13~kW. In the third hour, the price drops and the baselines fall to 72 and 77~kW. At these baselines, zero metered consumption gives a response in the partial-compensation tier of~\eqref{eq:dr_revenue}. The batteries recharge, raising the adjusted loads to 291 and 270~kW. The ESS discharges again during the evening peak from 15:00, and EVs depart fully charged.

Comfort is maintained between 23.0 and 23.8\,$^\circ$C on the illustrated 48-h trajectory, as shown in Fig.~\ref{fig:soc_ev}(a). SAC, DDPG-Lag, TD3 and DDPG leave the comfort band in 7, 4, 2 and 2 of these hours, respectively. The EV SoC at the 150 departures of each DRL method over the 5 seeds for the first test realisation is shown in Fig.~\ref{fig:soc_ev}(b). SAC, TD3 and DDPG leave 65, 14 and 12 departures below $\text{SoC}_{\mathrm{lim}}$, respectively, whereas DDPG-Lag and the proposed method leave 3 and 2. Most departures under DDPG-Lag and the proposed method end above the target, providing a margin for early departure.

\subsection{Interpretability Analysis}

\begin{table}[tb]
\centering
\caption{Fidelity of the distilled trees}
\label{tab:fidelity}
\setlength{\tabcolsep}{3pt}
\begin{tabular}{lcccc}
\toprule
Tree & Leaves & $R^2$, held-out & \shortstack{Agreement,\\test month (\%)} & \shortstack{Corrected\\leaves} \\
\midrule
ESS & 22.4\,$\pm$\,5.5 & 0.91\,$\pm$\,0.04 & 91.2\,$\pm$\,4.4 & 1.8\,$\pm$\,2.2 \\
EV & 13.0\,$\pm$\,2.8 & 0.67\,$\pm$\,0.15 & 79.6\,$\pm$\,12.6 & 0.0\,$\pm$\,0.0 \\
$r_{\mathrm{dis}}$ & 3.6\,$\pm$\,3.6 & 0.35\,$\pm$\,0.58 & 94.5\,$\pm$\,7.5 & -- \\
HVAC & 5.8\,$\pm$\,1.1 & 0.45\,$\pm$\,0.11 & -- & -- \\
\bottomrule
\end{tabular}
\end{table}

\begin{table}[tb]
\centering
\caption{Rules of the trees distilled from DDPG-Lag}
\label{tab:rules}
\setlength{\tabcolsep}{3pt}
\renewcommand{\arraystretch}{1.1}
\begin{tabular}{l >{\raggedright\arraybackslash}p{4.3cm} r r}
\toprule
Tree & If (prices in RMB/kWh, temperatures in $^\circ$C) & Then & States \\
\midrule
\multirow{5}{*}{ESS} & 03:00 to 15:00, $h_{\mathrm{DR}} > 0$, $p \le 0.80$ & $+77$ kW & 39\% \\
 & 03:00 to 15:00, $h_{\mathrm{DR}} > 0$, $p > 0.80$ & $-44$ kW & 2\% \\
 & 03:00 to 15:00, $h_{\mathrm{DR}} = 0$, $L^{\mathrm{base}} > 55.5$ kW & $-94$ kW & 6\% \\
 & 03:00 to 15:00, $h_{\mathrm{DR}} = 0$, $L^{\mathrm{base}} \le 55.5$ kW & $+28$ kW & 3\% \\
 & 15:00 to 03:00 & $-89$ to $-83$ kW & 50\% \\
\addlinespace
\multirow{4}{*}{EV} & $h_{\mathrm{DR}} = 0$ & $-90$ to $+73$ kW & 20\% \\
 & $h_{\mathrm{DR}} > 0$, $h_{\mathrm{dep}} \le 6$ & $+93$ kW & 52\% \\
 & $h_{\mathrm{DR}} > 0$, $h_{\mathrm{dep}} > 6$, $p \le 0.80$ & $+95$ kW & 22\% \\
 & $h_{\mathrm{DR}} > 0$, $h_{\mathrm{dep}} > 6$, $p > 0.80$ & $-100$ kW & 5\% \\
\addlinespace
\multirow{5}{*}{HVAC} & Before 07:00 & $+6$ kW & 29\% \\
 & 07:00 to 10:00, $T_{\mathrm{in}} > 22.5$ & $-17$ kW & 12\% \\
 & 10:00 to 13:00, $T_{\mathrm{free}} \le 24.1$ & $-8$ kW & 5\% \\
 & 10:00 to 13:00, $T_{\mathrm{free}} > 24.1$ & $+5$ kW & 7\% \\
 & After 13:00 & $+11$ kW & 46\% \\
\addlinespace
\multirow{2}{*}{$r_{\mathrm{dis}}$} & $L^{\mathrm{pro}} > 31.3$ kW, $L^{\mathrm{base}} > 14.8$ kW, $P^{\mathrm{PV}} > 20.2$ kW & 1.00 & 83\% \\
 & Other DR hours & 0.02 to 0.91 & 17\% \\
\bottomrule
\end{tabular}
\end{table}

Grouped battery-dispatch and HVAC residual rules are extracted from DDPG-Lag, as summarised in Table~\ref{tab:rules} for the first seed. The state shares refer to the training data. Battery powers are positive for charging, and HVAC outputs are residual offsets to $P^{\mathrm{req}}_{\mathrm{HVAC}}$, with negative values denoting additional cooling.

The ESS tree splits first on the hour, while the EV tree splits first on the physics-informed time to DR. Their rules in Table~\ref{tab:rules} store energy before DR, discharge during DR and the evening peak, and prioritise EV charging near departure.

The HVAC rules specify 17~kW of additional cooling before the DR window and 8~kW during DR while $T_{\mathrm{free}}\le24.1\,^\circ\mathrm{C}$. During cooling, projection prevents positive offsets from cancelling the required command. The allocation tree specifies $r_{\mathrm{dis}}=1$ in a region covering 83\% of its training DR states. Across seeds, $h_{\mathrm{DR}}$ contributes 31.4\% of EV-tree error reduction and $T_{\mathrm{free}}$ contributes 23.1\% for HVAC.

The proposed deployment is stable across tree depths from 2 to 8. Each seed's maximum cost is within 1.3\% of its minimum, with 0.04~h of average comfort violations and 0.6 departure violations throughout. Mean decision time is 0.03~ms. The fully tree-based ablation has a mean cost of 35\,208~RMB and the same comfort. Its within-seed cost range reaches 18.3\%, with up to 4.2 departure violations and DR compensation rates of 58\% to 61\% in individual seeds. The hybrid controller stabilises intertemporal battery scheduling across tree depths.

The battery-tree rules reproduce DDPG-Lag's command modes with 5-seed mean agreement of 91.2\% for the ESS and 79.6\% for EVs, as reported in Table~\ref{tab:fidelity}. These scores use all hours for the ESS and on-site hours for EVs in the first test realisation, with an idle band of $\pm$10\% of rated power. Held-out $R^2$ uses DDPG-Lag action or HVAC residual targets before the leaf check, while command agreement uses the checked trees. The allocation ratio differs from DDPG-Lag by at most 0.1 in 94.5\% of DR hours. HVAC is assessed through closed-loop comfort and cost.

\subsection{Ablation Study}

DDPG-Lag improves departure compliance relative to both smaller-step variants, as shown in the lower block of Table~\ref{tab:performance_comparison}. The fixed variant uses a step size of 0.005, and the adaptive variant increases it with mean violation up to 0.05 for comfort and 0.02 for departures. Both use zero tolerance and average SoC deficits over violated departures. DDPG corresponds to zero dual step. With the proposed update, departure violations decrease from 3.2 and 2.0 to 0.6, and cost decreases by 12.6\% relative to the fixed variant, with a p-value of 0.020.

Dynamic allocation and ageing costs lower operating cost under the small adaptive-step update. Fixing $r_{\mathrm{dis}} = 0.5$ raises the mean net operating cost from 36\,270 to 38\,828~RMB, by 7.1\%, because half of the battery discharge is then credited to the office zone and the DR response rate falls from 77.6\% to 70.9\%. Removing the ageing cost from training raises the mean cost by 12.2\% to 40\,687~RMB and nearly triples its spread across the seeds.

\subsection{Sensitivity Analysis}

Operating cost varies by less than 2\% across forecast-error standard deviations from 0 to 2\,$^\circ$C. Over this range, the proposed method records 0 to 11.9~h of comfort violations, with its maximum below the 17.0~h of the rule-based strategy and 49.1~h of DDPG-Lag. With EV departures up to 2~h early, it records 3.4 departure violations, below the 8.2 to 11.0 of SAC, TD3 and DDPG. For actual HVAC gains 10\% and 20\% below the nominal model, comfort violations are 3.1 and 43.2~h, below the 8.8 and 53.2~h of the rule-based strategy and 56.5 and 110.6~h of DDPG-Lag. All policies are evaluated without retraining.

\section{Conclusion}\label{sec:Conclusion}

In this article, we presented an interpretable constrained DRL method for industrial-park energy management with PV, heterogeneous batteries and production and office loads under tiered DR. DDPG-Lag learns the dispatch through an augmented Lagrangian reward with dual-ascent multipliers for comfort and EV departure constraints. Physics-informed distillation converts the policy into shallow regression trees. At deployment, office temperature is controlled by the residual HVAC tree with the model-predicted comfort projection. DDPG-Lag determines battery dispatch, and the battery trees provide readable scheduling rules. The distilled rules describe off-peak ESS charging, battery discharge during DR and the evening peak, and EV charging before departure. Relative to DRL methods with fixed penalty weights, comfort violations were reduced by 99.9\% and EV departure violations by 72.7\% to 95.4\%. Case studies used operational data from an industrial park in Fujian Province, China, with outdoor temperature forecast errors and uncertain EV departures. The proposed method achieved 0.04~h of average comfort violations and 98\% departure compliance, while reducing net operating cost by 9.5\% relative to a rule-based strategy.


\begin{thebibliography}{10}
\providecommand{\url}[1]{#1}
\csname url@samestyle\endcsname
\providecommand{\newblock}{\relax}
\providecommand{\bibinfo}[2]{#2}
\providecommand{\BIBentrySTDinterwordspacing}{\spaceskip=0pt\relax}
\providecommand{\BIBentryALTinterwordstretchfactor}{4}
\providecommand{\BIBentryALTinterwordspacing}{\spaceskip=\fontdimen2\font plus
\BIBentryALTinterwordstretchfactor\fontdimen3\font minus
  \fontdimen4\font\relax}
\providecommand{\BIBforeignlanguage}[2]{{\expandafter\ifx\csname l@#1\endcsname\relax
\typeout{** WARNING: IEEEtran.bst: No hyphenation pattern has been}\typeout{** loaded for the language `#1'. Using the pattern for}\typeout{** the default language instead.}\else
\language=\csname l@#1\endcsname
\fi
#2}}
\providecommand{\BIBdecl}{\relax}
\BIBdecl

\bibitem{datta2021review}
U.~Datta, A.~Kalam, and J.~Shi, ``A review of key functionalities of battery
  energy storage system in renewable energy integrated power systems,''
  \emph{Energy Storage}, vol.~3, no.~5, p. e224, 2021.

\bibitem{hannan2021battery}
M.~A. Hannan, S.~B. Wali, P.~J. Ker, M.~S. Abd~Rahman, M.~Mansor, V.~K.
  Ramachandaramurthy, K.~M. Muttaqi, T.~M.~I. Mahlia, and Z.~Y. Dong, ``Battery
  energy-storage system: A review of technologies, optimization objectives,
  constraints, approaches, and outstanding issues,'' \emph{J. Energy Storage},
  vol.~42, p. 103023, 2021.

\bibitem{yilmaz2012review}
M.~Yilmaz and P.~T. Krein, ``Review of benefits and challenges of
  vehicle-to-grid technology,'' in \emph{Proc. IEEE Energy Convers. Congr.
  Expo. (ECCE)}, 2012, pp. 3082--3089.

\bibitem{zhang2024multi}
X.~Zhang, Y.~Liu, Y.~Zhan, R.~Yan, J.~Mei, A.~Fu, F.~Jiao, and R.~Zeng,
  ``Multi-scenario optimization and performance evaluation of integrated energy
  system considering co-scheduling of {EVs} and stationary energy storage,''
  \emph{Renew. Energy}, vol. 237, p. 121927, 2024.

\bibitem{liu2026deep}
C.~Liu, Z.~Xu, J.~Zhou, Y.~Yuan, K.~Ma, and M.~Yuan, ``Deep reinforcement
  learning-based energy scheduling for green buildings with stationary and {EV}
  batteries of heterogeneous characteristics,'' \emph{Appl. Energy}, vol. 409,
  p. 127463, 2026.

\bibitem{siano2014demand}
P.~Siano, ``Demand response and smart grids---{A} survey,'' \emph{Renew.
  Sustain. Energy Rev.}, vol.~30, pp. 461--478, 2014.

\bibitem{dey2023microgrid}
B.~Dey, S.~Misra, and F.~P. Garcia~Marquez, ``Microgrid system energy
  management with demand response program for clean and economical operation,''
  \emph{Appl. Energy}, vol. 334, p. 120717, 2023.

\bibitem{tang2019model}
R.~Tang and S.~Wang, ``Model predictive control for thermal energy storage and
  thermal comfort optimization of building demand response in smart grids,''
  \emph{Appl. Energy}, vol. 242, pp. 873--882, 2019.

\bibitem{2004Benefits}
T.~A. Reddy, J.~K. Lukes, and L.~K. Norford, ``Benefits of multi-building
  electric load aggregation: {Actual} and simulation case studies,''
  \emph{ASHRAE Trans.}, vol. 110, no.~2, pp. 130--144, 2004.

\bibitem{2017Conceptual}
X.~Zhang, M.~Pipattanasomporn, M.~Kuzlu, and S.~Rahman, ``Conceptual framework
  for a multi-building peak load management system,'' in \emph{Proc. IEEE PES
  Innov. Smart Grid Technol. Conf. Europe (ISGT-Europe)}, 2016, pp. 1--5.

\bibitem{yuan2026reinforcement}
M.~Yuan, Y.~Wang, X.~Yu, T.~Wik, and C.~Zou, ``Reinforcement learning-based
  home energy management with heterogeneous batteries and stochastic {EV}
  behaviour,'' \emph{IEEE Trans. Sustain. Energy}, 2026, {Early Access}, doi:
  10.1109/TSTE.2026.3719940.

\bibitem{selvaraj2023smart}
R.~Selvaraj, V.~M. Kuthadi, and S.~Baskar, ``Smart building energy management
  and monitoring system based on artificial intelligence in smart city,''
  \emph{Sustain. Energy Technol. Assess.}, vol.~56, p. 103090, 2023.

\bibitem{thirugnanam2022energy}
K.~Thirugnanam, M.~S. El~Moursi, V.~Khadkikar, H.~H. Zeineldin, and
  M.~Al~Hosani, ``Energy management strategy of a reconfigurable grid-tied
  hybrid {AC/DC} microgrid for commercial building applications,'' \emph{IEEE
  Trans. Smart Grid}, vol.~13, no.~3, pp. 1720--1738, 2022.

\bibitem{schmitt2023regression}
T.~Schmitt, J.~Engel, and T.~Rodemann, ``Regression-based model error
  compensation for a hierarchical {MPC} building energy management system,'' in
  \emph{Proc. IEEE Conf. Control Technol. Appl. (CCTA)}, 2023, pp. 1--8.

\bibitem{khan2022modeling}
M.~H. Khan, A.~U. Asar, N.~Ullah, F.~R. Albogamy, and M.~K. Rafique, ``Modeling
  and optimization of smart building energy management system considering both
  electrical and thermal load,'' \emph{Energies}, vol.~15, no.~2, p. 574, 2022.

\bibitem{kuboth2019economic}
S.~Kuboth, F.~Heberle, A.~K{\"o}nig-Haagen, and D.~Br{\"u}ggemann, ``Economic
  model predictive control of combined thermal and electric residential
  building energy systems,'' \emph{Appl. Energy}, vol. 240, pp. 372--385, 2019.

\bibitem{cao2020deep}
J.~Cao, D.~Harrold, Z.~Fan, T.~Morstyn, D.~Healey, and K.~Li, ``Deep
  reinforcement learning-based energy storage arbitrage with accurate
  lithium-ion battery degradation model,'' \emph{IEEE Trans. Smart Grid},
  vol.~11, no.~5, pp. 4513--4521, 2020.

\bibitem{2024Demand}
E.~J. Salazar, V.~Rosero, J.~Gabrielski, and M.~E. Samper, ``Demand response
  model: {A} cooperative-competitive multi-agent reinforcement learning
  approach,'' \emph{Eng. Appl. Artif. Intell.}, vol. 133, p. 108273, 2024.

\bibitem{pei2025hierarchical}
Y.~Pei, Y.~Yao, J.~Zhao, J.~Hao, F.~Ding, and J.~Wang, ``Multi-agent
  hierarchical deep reinforcement learning for {HVAC} control with flexible
  {DERs},'' \emph{IEEE Trans. Smart Grid}, vol.~16, no.~6, pp. 5589--5601,
  2025.

\bibitem{su2025review}
T.~Su, T.~Wu, J.~Zhao, A.~Scaglione, and L.~Xie, ``A review of safe
  reinforcement learning methods for modern power systems,'' \emph{Proc. IEEE},
  vol. 113, no.~3, pp. 213--255, 2025.

\bibitem{altman1999constrained}
E.~Altman, \emph{Constrained {Markov} Decision Processes}.\hskip 1em plus 0.5em
  minus 0.4em\relax Boca Raton, FL, USA: Chapman \& Hall/CRC, 1999.

\bibitem{achiam2017constrained}
J.~Achiam, D.~Held, A.~Tamar, and P.~Abbeel, ``Constrained policy
  optimization,'' in \emph{Proc. 34th Int. Conf. Mach. Learn. (ICML)}, 2017,
  pp. 22--31.

\bibitem{tessler2019reward}
C.~Tessler, D.~J. Mankowitz, and S.~Mannor, ``Reward constrained policy
  optimization,'' in \emph{Proc. 7th Int. Conf. Learn. Represent. (ICLR)},
  2019.

\bibitem{chen2023interpretable}
Z.~Chen, F.~Xiao, F.~Guo, and J.~Yan, ``Interpretable machine learning for
  building energy management: {A} state-of-the-art review,'' \emph{Adv. Appl.
  Energy}, vol.~9, p. 100123, 2023.

\bibitem{rudin2019stop}
C.~Rudin, ``Stop explaining black box machine learning models for high stakes
  decisions and use interpretable models instead,'' \emph{Nat. Mach. Intell.},
  vol.~1, no.~5, pp. 206--215, 2019.

\bibitem{bastani2018verifiable}
O.~Bastani, Y.~Pu, and A.~Solar-Lezama, ``Verifiable reinforcement learning via
  policy extraction,'' in \emph{Proc. Adv. Neural Inf. Process. Syst.
  (NeurIPS)}, vol.~31, 2018.

\bibitem{chen2025novel}
H.~Chen, W.~Guo, W.~Bao, M.~Cui, X.~Wang, and Q.~Zhao, ``A novel interpretable
  decision rule extracting method for deep reinforcement learning-based energy
  management in building complexes,'' \emph{Energy Build.}, vol. 334, p.
  115514, 2025.

\bibitem{lu2020optimal}
Q.~Lu, S.~L{\"u}, Y.~Leng, and Z.~Zhang, ``Optimal household energy management
  based on smart residential energy hub considering uncertain behaviors,''
  \emph{Energy}, vol. 195, p. 117052, 2020.

\bibitem{fujian2024demandresponse}
{Fujian Provincial Development and Reform Commission}, ``Fujian electric power
  demand response implementation plan,''
  \url{https://fgw.fj.gov.cn/zfxxgkzl/zfxxgkml/bwgfxwj/202407/t20240723_6487310.htm},
  Jul. 2024.

\bibitem{naumann2020analysis}
M.~Naumann, F.~B. Spingler, and A.~Jossen, ``Analysis and modeling of cycle
  aging of a commercial {LiFePO$_4$}/graphite cell,'' \emph{J. Power Sources},
  vol. 451, p. 227666, 2020.

\bibitem{schmalstieg2014holistic}
J.~Schmalstieg, S.~K{\"a}bitz, M.~Ecker, and D.~U. Sauer, ``A holistic aging
  model for {Li(NiMnCo)O$_2$} based 18650 lithium-ion batteries,'' \emph{J.
  Power Sources}, vol. 257, pp. 325--334, 2014.

\bibitem{lillicrap2016continuous}
T.~P. Lillicrap, J.~J. Hunt, A.~Pritzel, N.~Heess, T.~Erez, Y.~Tassa,
  D.~Silver, and D.~Wierstra, ``Continuous control with deep reinforcement
  learning,'' in \emph{Proc. 4th Int. Conf. Learn. Represent. (ICLR)}, 2016.

\bibitem{breiman1984classification}
L.~Breiman, J.~H. Friedman, R.~A. Olshen, and C.~J. Stone, \emph{Classification
  and Regression Trees}.\hskip 1em plus 0.5em minus 0.4em\relax Belmont, CA,
  USA: Wadsworth, 1984.

\bibitem{ross2011reduction}
S.~Ross, G.~J. Gordon, and J.~A. Bagnell, ``A reduction of imitation learning
  and structured prediction to no-regret online learning,'' in \emph{Proc. 14th
  Int. Conf. Artif. Intell. Statist. (AISTATS)}, 2011, pp. 627--635.

\bibitem{bnef2025battery}
{BloombergNEF}, ``Lithium-ion battery pack prices fall to \$108 per
  kilowatt-hour, despite rising metal prices: {BloombergNEF},''
  \url{https://about.bnef.com/insights/clean-transport/lithium-ion-battery-pack-prices-fall-to-108-per-kilowatt-hour-despite-rising-metal-prices-bloombergnef/},
  2025, press release, {Dec.} 9, 2025. {Accessed}: {Oct.} 7, 2026.

\bibitem{fujimoto2018addressing}
S.~Fujimoto, H.~van Hoof, and D.~Meger, ``Addressing function approximation
  error in actor-critic methods,'' in \emph{Proc. 35th Int. Conf. Mach. Learn.
  (ICML)}, 2018, pp. 1587--1596.

\bibitem{haarnoja2018soft}
T.~Haarnoja, A.~Zhou, P.~Abbeel, and S.~Levine, ``Soft actor-critic:
  {Off}-policy maximum entropy deep reinforcement learning with a stochastic
  actor,'' in \emph{Proc. 35th Int. Conf. Mach. Learn. (ICML)}, 2018, pp.
  1861--1870.

\end{thebibliography}
\end{document}